\documentclass[conference]{IEEEtran}
\IEEEoverridecommandlockouts
\usepackage{cite}
\usepackage{amsmath,amssymb,amsfonts}
\usepackage{algorithmic}
\usepackage{graphicx}
\usepackage{textcomp}
\usepackage{xcolor}
\usepackage{caption}
\usepackage{subcaption}
\usepackage{multirow}
\usepackage{booktabs}
\usepackage{todonotes}
\def\BibTeX{{\rm B\kern-.05em{\sc i\kern-.025em b}\kern-.08em
    T\kern-.1667em\lower.7ex\hbox{E}\kern-.125emX}}
\begin{document}

\title{Edge-Aware Autoencoder Design for Real-Time Mixture-of-Experts Image
Compression}

\author{\IEEEauthorblockN{Elvira Fleig, Jonas Geistert, Erik Bochinski, Rolf Jongebloed, and Thomas Sikora}
\IEEEauthorblockA{\textit{Communication Systems Group} \\
\textit{Technische Universität Berlin}\\
Berlin, Germany}
}

\maketitle

\begin{abstract} 
Steered-Mixtures-of-Experts (SMoE) models provide sparse, edge-aware representations, applicable to many use-cases in image processing. This includes denoising, super-resolution and compression of 2D- and higher dimensional pixel data. Recent works for image compression indicate that compression of images based on SMoE models can provide competitive performance to the state-of-the-art. Unfortunately, the iterative model-building process at the encoder comes with excessive computational demands.\\
In this paper we introduce a novel edge-aware Autoencoder (AE) strategy designed to avoid the time-consuming iterative optimization of SMoE models. This is done by directly mapping pixel blocks to model parameters for compression, in spirit similar to recent works on “unfolding” of algorithms, while maintaining full compatibility to the established SMoE framework. With our plug-in AE encoder, we achieve a quantum-leap in performance with encoder run-time savings by a factor of 500 to 1000 with even improved image reconstruction quality. For image compression the plug-in AE encoder has real-time properties and improves RD-performance compared to our previous works.
\end{abstract}


\section{Introduction} 

The Steered Mixture-of-Experts (SMoE) approach has been first presented as a promising regression framework for coding images \cite{verhack2016universal}\cite{jongebloed2018hierarchical}\cite{liu2019image}\cite{bochinski2018regularized}\cite{tok2018mse} and has since expanded to other application fields such as video compression \cite{lange2016video} and coding of higher dimensional data, like light fields and light field videos \cite{verhack2019steered}. 

The intriguing property of the SMoE approach is its capability to model in-stationarities in images (such as edges or smooth transitions) efficiently with a sparse edge-aware representation. Swarms of steered kernels are employed to model image pixel properties.
Once the model is optimized during encoding, all kernels are aligned to harvest best possible correlation between pixels – and afterwards collaborate for reconstruction of the pixel amplitudes at the decoder.
Compression of images is performed in the image-domain rather than in a transform domain by quantizing the SMoE kernel parameters.

Model parameters are optimized at the encoder network for each block to be coded, i.e. using Maximum-Expectation or Gradient-Descent algorithms. 
For best model quality the Gradient-Descent optimization is preferred \cite{bochinski2018regularized}\cite{jongebloed2019quantized}\cite{tok2018mse}.
An advantage of this approach is, that cost functions such as mean-squared-error or SSIM with flexible side conditions can be used. For quantization of the model parameters, strategies like “fake quantization” can be readily employed within the optimization process to approach Rate-Distortion bounds. 
The reader is referred to the work by Jongebloed et al. \cite{jongebloed2019quantized} for competitive compression results of SMoE compared to JPEG2000.
Unfortunately, to achieve reasonable model quality, many training iterations are required which leads to a high computational cost. This makes the SMoE coding approach unfeasible for real-time applications.

In this paper we present a novel approach for estimating SMoE parameters via the design of a novel Autoencoder network with embedded SMoE decoder capabilities.
The strategy we are suggesting provides a break-through in run-time performance with essentially no compromise to the quality of the reconstructed images.\\
To show the benefits of our strategy for compression we investigate our novel SMoE Autoencoder in the context of the SMoE image coding framework of Tok et al. in \cite{tok2018mse}. 
As we shall see, we are able to achieve real-time capabilities with encoder run-time savings by gains of 500-1000 compared to Tok et al. without sacrificing rate-distortion performance. 
\section{The Sparse, edge-aware SMoE image model}
\begin{figure}
\begin{subfigure}[t]{0.24\linewidth}
\includegraphics[width=\linewidth]{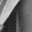}
\caption{Original}
\label{fig:1_orig}
\end{subfigure}
\hfill
\begin{subfigure}[t]{0.24\linewidth}
\includegraphics[width=\linewidth]{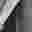}
\caption{\textbf{JPEG}\\ PSNR:$26.33 dB$\\ SSIM: $0.82$}
\label{fig:1_JPEG}
\end{subfigure}%
\hfill
\begin{subfigure}[t]{0.24\linewidth}
\includegraphics[width=\linewidth]{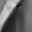}
\caption{\textbf{HEVC}\\PSNR:$26.05 dB$ \\SSIM: $0.77$}
\label{fig:1_HEVC}
\end{subfigure}%
\hfill
\begin{subfigure}[t]{0.24\linewidth}
\includegraphics[width=\linewidth]{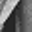}
\caption{\textbf{JPEG2000}\\ PSNR:$29.43 dB$\\ SSIM: $0.87$}
\label{fig:1_JPEG200}
\end{subfigure}%
\hfill
\begin{subfigure}[t]{0.24\linewidth}
\includegraphics[width=\linewidth]{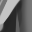}
\caption{\textbf{SMoE}\\ PSNR:$31.66 dB$ \\SSIM: $0.9$}
\label{fig:1_smoe}
\end{subfigure}%
\hfill
\begin{subfigure}[t]{0.24\linewidth}
\includegraphics[width=\linewidth]{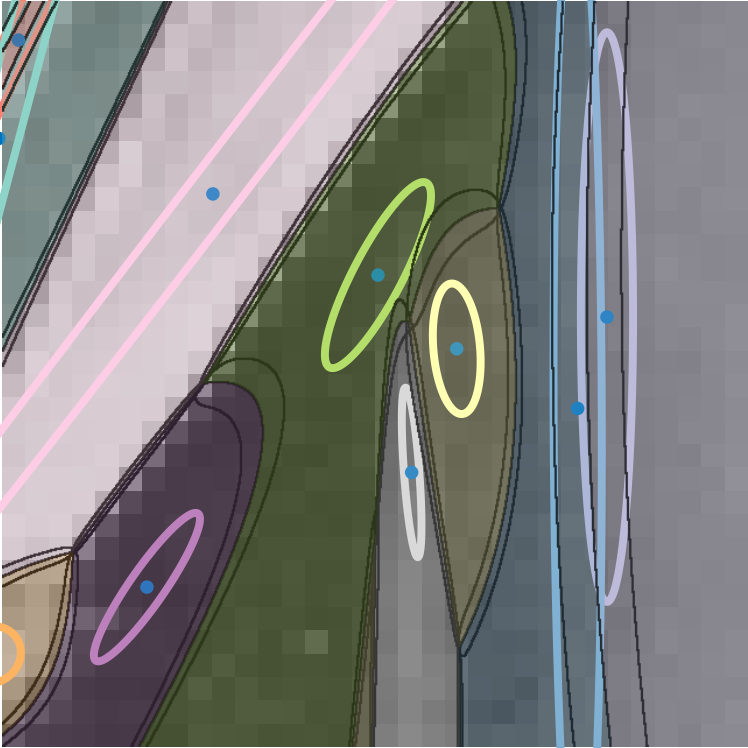}
\caption{SMoE Kernels}
\label{fig:1_kernels}
\end{subfigure}%
\hfill
\begin{subfigure}[t]{0.24\linewidth}
\includegraphics[width=\linewidth]{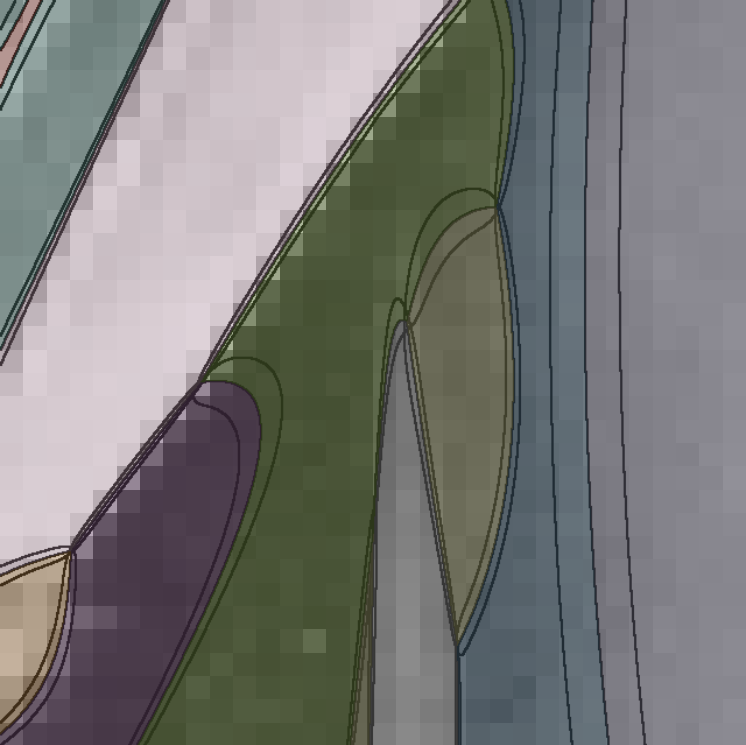}
\caption{SMoE Gates}
\label{fig:1_Gates}
\end{subfigure}%
\caption{Compression performance at 0.43 bpp}
\label{fig:Comparison_JPEG_HEVC_SMoE}
\vspace{-0.2cm}
\end{figure}

The SMoE image model describes an edge-aware, parametric, continuous non-linear regression function. We use Fig. \ref{fig:Comparison_JPEG_HEVC_SMoE} to illustrate the kernel model concept and the capabilities of the model to reconstruct sharp and smooth transitions in images without common ringing artifacts seen in JPEG-like coders. The sparse edge-aware SMoE model is able to reconstruct the $32\times32$ pixel block with excellent edge quality. In contrast to JPEG, JPEG2000 and HEVC-Intra at same bit rate, typical blocking and ringing artifacts are avoided completely. Both objective quality measures as well as subjective quality are greatly enhanced. The locations and steering properties of the gaussian kernels after optimization explain the correlation between pixels. The resulting 2D-gating functions provide for the edge-awareness of the model. Sharp edges are modelled with sharp gating functions while smooth transitions are modelled using overlapping gates. The reader is referred to \cite{jongebloed2018hierarchical}\cite{verhack2019steered}\cite{jongebloed2019quantized} for more insight into the subject.
This edge-aware sparse representation of SMoE’s appears to be attractive for a number of reasons for compression and beyond:
\begin{itemize}
    \item SMoE’s easily extend to N-dimensional signals, including light-fields and beyond \cite{verhack2017steered}.
    \item Modeling is not restricted to regular N-dimensional pixel grids, but can be applied to irregularly sampled imagery including arbitrarily shaped segments and point clouds of any dimension.
    \item The sparse (coded) representation allows to easily extract N-dimensional image features from the (decoded) model, such as edges, intensity flow, etc.\cite{verhack2016universal}.
    \item Once built, the continuous SMoE model allows to resample to any resolution in time and space, which includes edge-aware super-resolution and motion-interpolation. 
\end{itemize}

\section{Theoretical Background}
Steered Mixtures-of-Experts regress the luminance value at each position $\underline{x}$ of an image by a weighted sum of $K$ experts: 
\begin{align}
    y_p(\underline{x})=\sum_{i=1}^{K} m_i(\underline{x}) \cdot w_i(\underline{x})
    \label{eq:Grundgleichung}
\end{align}
with $m_i(\underline{x})$ being the experts and $w_i(\underline{x})$ are gating functions.
In this work, we focus on constant experts as of Eq. \ref{eq:constexp}. However, other functions like hyperplanes \cite{verhack2016universal} or polynomials can be used instead.
\begin{align}
    f_i(\underline{x})=m_i
    \label{eq:constexp}
\end{align}
Respectively the gating functions $w_i(\underline{x})$ are defined as weighted soft-max functions
\begin{align}
    w_i(\underline{x})=\frac{\mathcal{K}(\underline{x},\underline{\mu}_i,\underline{\Sigma}_i)}{\sum_{j=1}^{K}\mathcal{K}(\underline{x},\underline{\mu}_j,\underline{\Sigma}_j)}
\end{align}
described by kernel functions 
$\mathcal{K}(\underline{x},\underline{\mu_i},\underline{\Sigma}_i)$, such as laplacians or as is the case in this work, gaussian kernels:
\begin{align}
   \mathcal{K}(\underline{x},\underline{\mu}_i,\underline{\Sigma}_i)=exp[-\frac{1}{2}(\underline{x}-\underline{\mu}_i)^T \underline{\Sigma}_i^{-1} (\underline{x}-\underline{\mu}_i)]
\end{align}
With $\underline{\mu}_i$ the center position of the kernels and $\underline{\Sigma}_i$ the steering parameters. 
This work uses a simplified definition by having radial kernels that do not steer in any direction and using kernels of the same bandwidth $S$.
So the regression function in Eq \ref{eq:Grundgleichung} simplifies to:
\begin{align}
   y(\underline{x})=\sum_{i=1}^{K} m_i\cdot\frac{exp(-\mathcal{S}||\underline{x}-\underline{\mu}_i||^2)}{\sum_{j=1}^{K}exp(-\mathcal{S}||\underline{x}-\underline{\mu}_j||^2)}
   \label{eq:OPGleichung}
\end{align}
Even with these simplifications, a good representation of smooth and sharp transitions can be achieved by adjusting the center positions of the kernels.
The possibility to model complex patterns increases with the number of used kernels.
The	work	of \cite{tok2018mse}	introduced	the gradient gescent	optimization	(SMoE-GD)	to	achieve	better
results	in	comparison	to	the EM	Algorithm. The	optimization	objective	is	to	directly	minimize	the	
reconstruction	error	instead	of	maximizing the joint likelihood of the location and pixel amplitude \cite{verhack2016universal} which	is	less	sensitive	to	subjective	image	
quality.
The	function	$\mathcal{L}$	is	defined	as	loss	to	perform	the mean-squared-error	optimization on all $N$ pixels in a block:	
\begin{align}
   \mathcal{L}:=\frac{1}{N}\sum_{n=1}^N(y_n(\underline{x})-\sum_{i=1}^K m_i(\underline{x}_n)\cdot w_i(\underline{x}_n))^2
\end{align}

\begin{align}
    \underset{m_i,\underline{\mu}_i}{\mathrm{arg\,min}} \{\mathcal{L}\}
\end{align}
This	approach was	used	in	Fig. \ref{fig:Comparison_JPEG_HEVC_SMoE} to	optimize	the	SMoE	model.	Its quality	is highly	dependable	on	the	initialization	of	the	kernels and	can	quickly	get	stuck	in	local	minima.
In	addition,	many	training iterations	are	necessary	to	achieve	a	good	quality	reconstruction,	as	will be discussed below.	This makes	the	iterative	approach	unfeasible	for	real-time applications.	

\section{Edge-Aware Autoencoder Design with Embedded SMoE Decoder}
\begin{figure}
\centering
\includegraphics[width=0.9\linewidth]{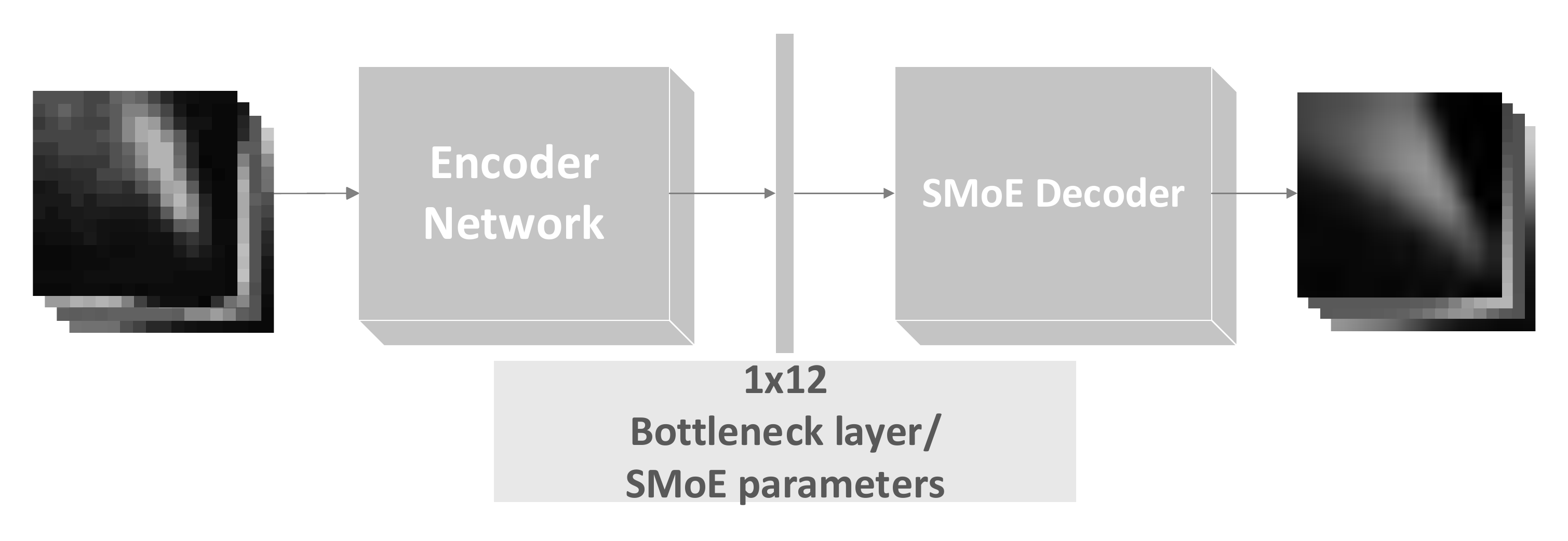}
\caption{SMoE Autoencoder}
\label{fig:structure_AE}
\vspace{-0.3cm}
\end{figure}

\subsection{Autoencoder Design Principles}
In order to circumvent the above drawbacks of the gradient descent optimization we propose to design a SMoE Autoencoder network (SMoE-AE) that generates SMoE parameters at the bottleneck layer ready for compression and coding.
The basic concept of the SMoE-AE network with embedded SMoE decoder is depicted in Fig. \ref{fig:structure_AE}. 
As such, we propose to design a deep encoder network in combination with a fixed, shallow SMoE decoder.\\
Once the SMoE-AE network parameters are optimized in a training process based on ground-truth imagery the SMoE-AE encoder network is fixed - and can be used to predict SMoE parameters for pixels of individual input blocks. 
In this way the above mentioned block-wise gradient-descent SMoE model building process develops into a “Learnt Compression” scenario.\\
It is our hypothesis that SMoE-AE networks with suitable network sizes and restricted number of network parameters can produce meaningful SMoE parameters of sufficient quality - with a processing time significantly less than those produced with gradient descent optimization. 
The envisioned SMoE-AE network is novel in concept because the fixed SMoE model is an integral part of the SMoE-AE end-to-end training process.
This ensures that the AE encoder network produces meaningful bottleneck SMoE parameters. 
It is readily feasible to design the encoders such that the SMoE parameters are ready for coding - discrete and quantized to approach RD-bounds. 
Any objective measure usually used in neural network learnt-compression approaches (such as MSE, SSIM and beyond) can be used for Rate-Distortion optimization of the SMoE-AE encoders. \\
In this paper we first present results with a bottleneck layer restricted to produce parameters for the radial SMoE decoder according to Eq. \ref{eq:OPGleichung} with $K=4$ radial kernels/block and bandwidth $S$ as hyper parameter. This makes it possible to compare our results directly with those of \cite{tok2018mse}. To comply with \cite{tok2018mse}, the autoencoder is built to provide parameters for gray-level pixels in blocks of size $16\times16$.\\
The SMoE-AE encoder is designed to follow the concept of conventional Autoencoders \cite{tschannen2018recent} with a $16\times16$ dimensional input and decreasing layer sizes towards the 12-dimensional latent space output.
For the decoder part, the shallow SMoE decoder network is used.
The architectural structure of the encoder is depicted in Tab. \ref{tab:Architecture}.
The last fully-connected layer presents the bottleneck of the autoencoder, and the values are then passed on to the SMoE decoder. 
The total number of parameters in the encoder network is 68,855,116. 
The innermost layer comprises the kernel center values $\underline{\mu}_i$ and the expert values $m_i$ to perform the SMoE reconstruction based on Eq.\ref{eq:OPGleichung}.\\
It appears obvious to evaluate and compare the proposed SMoE-AE with an conventional AE network (C-AE) with symmetric encoder-decoder design by replacing the shallow SMoE decoder network with the traditional “mirrored” decoder network.
The question then is how the C-AE network, given same size and topology and same size of bottleneck layer, compares to the performance of the edge-aware SMoE-AE. 
Such an AE network is optimized by jointly optimizing both AE encoder and decoder network parameters in an end-to-end fashion.
Intuitively one is tempted to think that the SMoE-AE with its fixed decoder (which is also shallow in contrast to the conventional AE decoder) cannot outperform the C-AE network which has twice the amount of trainable parameters. 
The SMoE-AE decoder network, however, models edges in images efficiently and it is our hypothesis that such models can help to derive AE networks with superior edge reconstruction quality.
\begin{table}
\begin{centering}

\begin{scriptsize}
\caption{Encoder architecture and training  parameter}
\label{tab:Architecture}
\centering
\begin{tabular}[c]{p{1cm} p{.6cm}  p{1.2cm} p{1.4cm} p{0.8cm} p{1.2cm}}
\toprule
\multicolumn{2}{c}{General}&\multicolumn{2}{c}{Conv. Layers}&\multicolumn{2}{c}{Dense Layer}\\
\midrule
Epochs&30&Quantity&6&Quantity&5\\
LR&0.00005&\multirow{2}{*}{Filter}&\multirow{2}{*}{\shortstack[l]{16, 32, 64,\\ 128, 256, 512}}&\multirow{2}{*}{\shortstack[l]{Output\\ Dim.}}&\multirow{2}{*}{\shortstack[l]{512, 265, 128,\\ 64, 12}}\\
Optimizer&Adam&&&&\\
Activation&ReLU&Padding&Same&&\\
Loss&MSE&Kernel Size&3x3&&\\
\bottomrule

\end{tabular}

\end{scriptsize}
\end{centering}
\vspace{-0.1cm}
\end{table}
\subsection{Training the SMoE-AE network parameters}
The mobile subset of the clic dataset \cite{CLIC2020} was used to train the networks. The test images used for coding in this paper were not part of the training set. All training images were converted to greyscale, cropped to $1024 \times 1024$ pixels and $16\times16$ pixel blocks were then extracted out of these images. Pixel values were scaled to the domain $[0 .. 1]$. The training set thus contained over four million individual blocks. 
For training we did not categorize blocks into textured or non-textured blocks as in \cite{tok2018mse}.
The order of the blocks were shuffled during training to ensure a suitable generalization of the network. Bottleneck parameters were also restricted to be in the domain $[0 .. 1]$. All training parameters are shown in Tab. \ref{tab:Architecture}. The training took about 20 hours on a Nvidia GTX 1070 Ti.
After training the encoder, the decoder part was decoupled and the encoder was used to predict the center positions $\underline{\mu}_i$ and expert values $m_i$ for each block. The C-AE encoder network was identical in size and topology to the SMoE encoder. Both C-AE encoder and decoder networks were optimized jointly which took about 25 hours.


\section{Experiments/Evaluation}
\begin{table}
\begin{scriptsize}

\caption{PSNR in [dB] and SSIM reconstruction quality}
\label{tab:Tok_Time}
\centering
\begin{tabular}[c]{l l p{.6cm} p{.6cm} p{.6cm} p{.6cm} p{.6cm} p{.6cm}  }
\toprule
Sequence&&\multicolumn{4}{c}{SMoE-GD}&\hspace*{-2mm}\mbox{SMoE-AE}& C-AE\\
\midrule
&Iter.&1&500&5000&20000&n/a&n/a\\
\midrule
\multirow{2}{*}{Cameraman}&PSNR&22.57&26.27&27.08&27.09&27.79&\textbf{27.92}\\
&SSIM&0.71&0.82&0.85&0.85&0.85&\textbf{0.85}\\

\multirow{2}{*}{Lena}&PSNR&24.06&27.82&28.83&28.91&\textbf{29.05}&28.87\\
&SSIM&0.65&0.79&0.82&0.82&\textbf{0.82}&0.81\\
\multirow{2}{*}{Peppers}&PSNR&23.84&28.43&29.5&29.55&\textbf{29.56}&29.36\\
&SSIM&0.62&0.76&0.79&\textbf{0.79}&0.78&0.78\\
\bottomrule

\end{tabular}
\end{scriptsize}
\vspace{-0.3cm}
\end{table}

\begin{figure}
\begin{subfigure}[t]{0.24\linewidth}
\includegraphics[width=\linewidth]{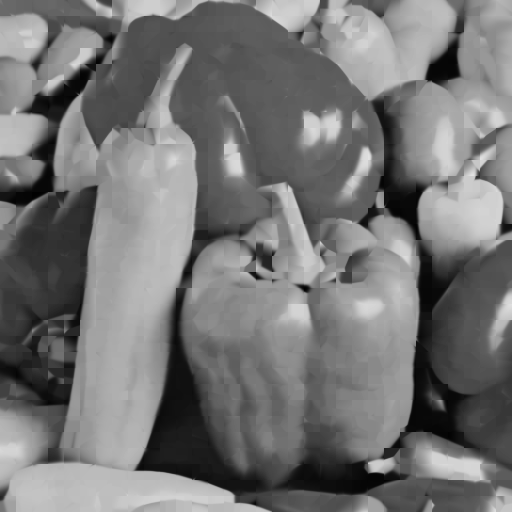}
\caption{SMoE-GD}
\label{fig:Peppers_GD_20k}
\end{subfigure}%
\hfill
\begin{subfigure}[t]{0.24\linewidth}
\includegraphics[width=\linewidth]{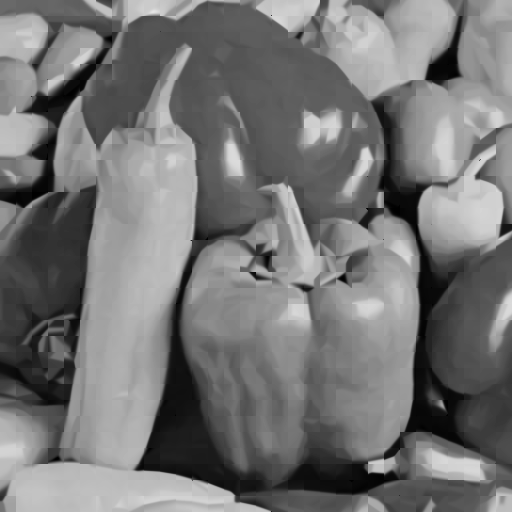}
\caption{SMoE-AE}
\label{fig:Peppers_SmoE-AE}
\end{subfigure}%
\hfill
\begin{subfigure}[t]{0.24\linewidth}
\includegraphics[width=\linewidth]{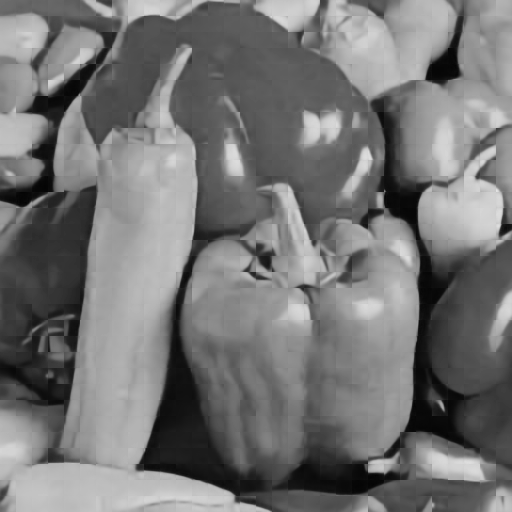}
\caption{C-AE}
\label{fig:Peppers_AE}
\end{subfigure}%
\caption{Reconstruction quality without quantization}
\label{fig:Comp_1_iteration}
\vspace{-0.2cm}
\end{figure} 

\begin{table}
\begin{centering}

\begin{scriptsize}
\caption{Encoder decoder run-time comparison}
\label{tab:Encode-Decode_Times}
\centering
\begin{tabular}[c]{l l l l l  }
\toprule
Time [s]&\multicolumn{2}{c}{SMoE-GD}& SMoE-AE& C-AE\\
\midrule
Iterations &5000&20000&n/a&n/a\\
\midrule
Encoding time & 296.29s &1191.22s&0.25s &0.25s\\
Decoding time & 0.02s & 0.02s & 0.02s & 0.25s\\

\bottomrule

\end{tabular}

\end{scriptsize}
\end{centering}

\end{table}

\begin{figure}

\begin{subfigure}[t]{0.25\linewidth}
\includegraphics[width=0.8\linewidth]{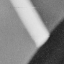}
\caption{\textbf{Original}}
\label{fig:Orig_Ausschnitt02}
\end{subfigure}%
\hfill
\begin{subfigure}[t]{0.25\linewidth}
\includegraphics[width=0.8\linewidth]{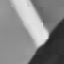}
\caption{\textbf{SMoE-AE} \\35.01 dB}
\label{fig:Smoe_Ausschnitt02}
\end{subfigure}%
\hfill
\begin{subfigure}[t]{0.25\linewidth}
\includegraphics[width=0.8\linewidth]{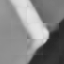}
\caption{\textbf{C-AE} \\32.47 dB}
\label{fig:AE_Ausschnitt02}
\end{subfigure}%
\caption{Edge-rconstruction for "Lena" crop} 
\label{fig:AE_Smoe_Ausschnitte}
\vspace{-0.3cm}
\end{figure}

\begin{figure}
\begin{centering}
\begin{subfigure}[t]{0.24\linewidth}
\includegraphics[width=\linewidth]{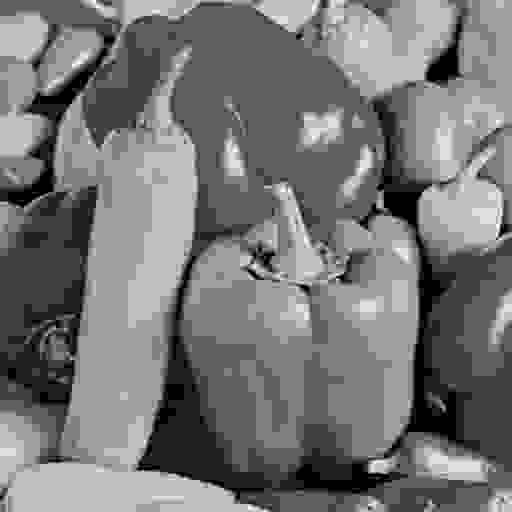}
\caption{\textbf{JPEG}\\ at 0.14bpp}
\label{fig:Peppers_JPEG_0-14}
\end{subfigure}
\begin{subfigure}[t]{0.24\linewidth}
\includegraphics[width=\linewidth]{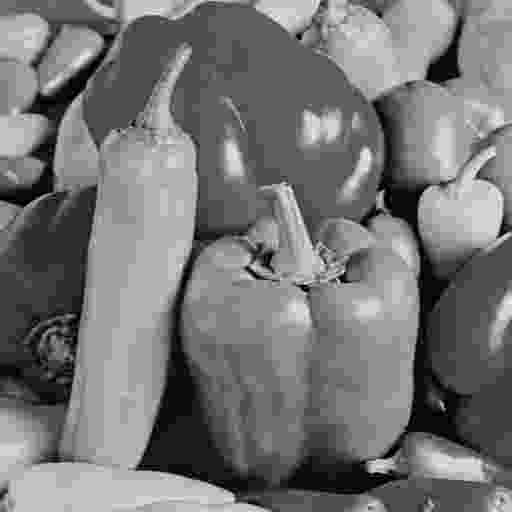}
\caption{\textbf{JPEG}\\at 0.18bpp}
\label{fig:Peppers_JPEG_0-18}
\end{subfigure}%
\hfill
\begin{subfigure}[t]{0.24\linewidth}
\includegraphics[width=\linewidth]{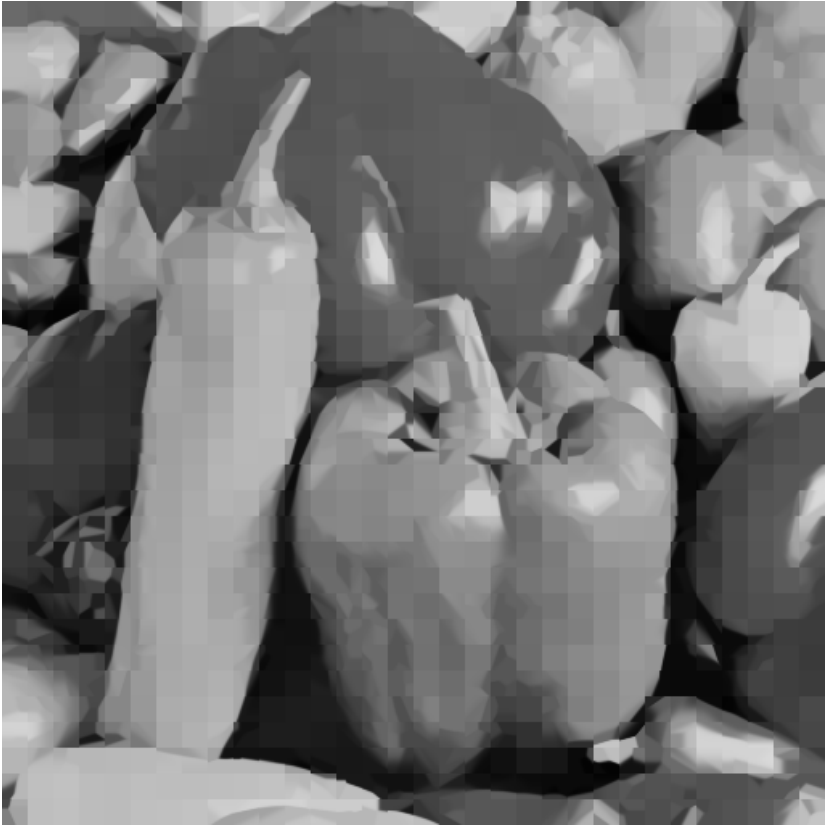}
\caption{\textbf{SMoE-GD}, at 0.14 bpp}
\label{fig:Peppers_SMoE-GD}
\end{subfigure}%
\hfill
\begin{subfigure}[t]{0.24\linewidth}
\includegraphics[width=\linewidth]{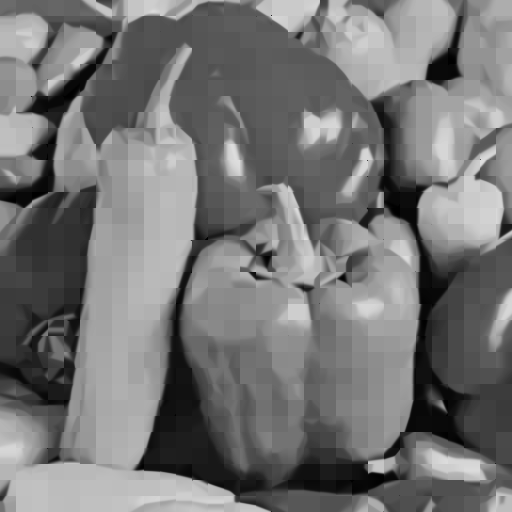}
\caption{\textbf{SMoE-AE} at 0.14 bpp}
\label{fig:Peppers_SMoE-AE}
\end{subfigure}%
\end{centering}
\caption{Comparison of JPEG and SMoE-AE}
\label{fig:JPEG_SMOE}
\vspace{-0.2cm}
\end{figure}

\begin{figure}
\begin{subfigure}[t]{0.49\linewidth}
\includegraphics[width=\linewidth]{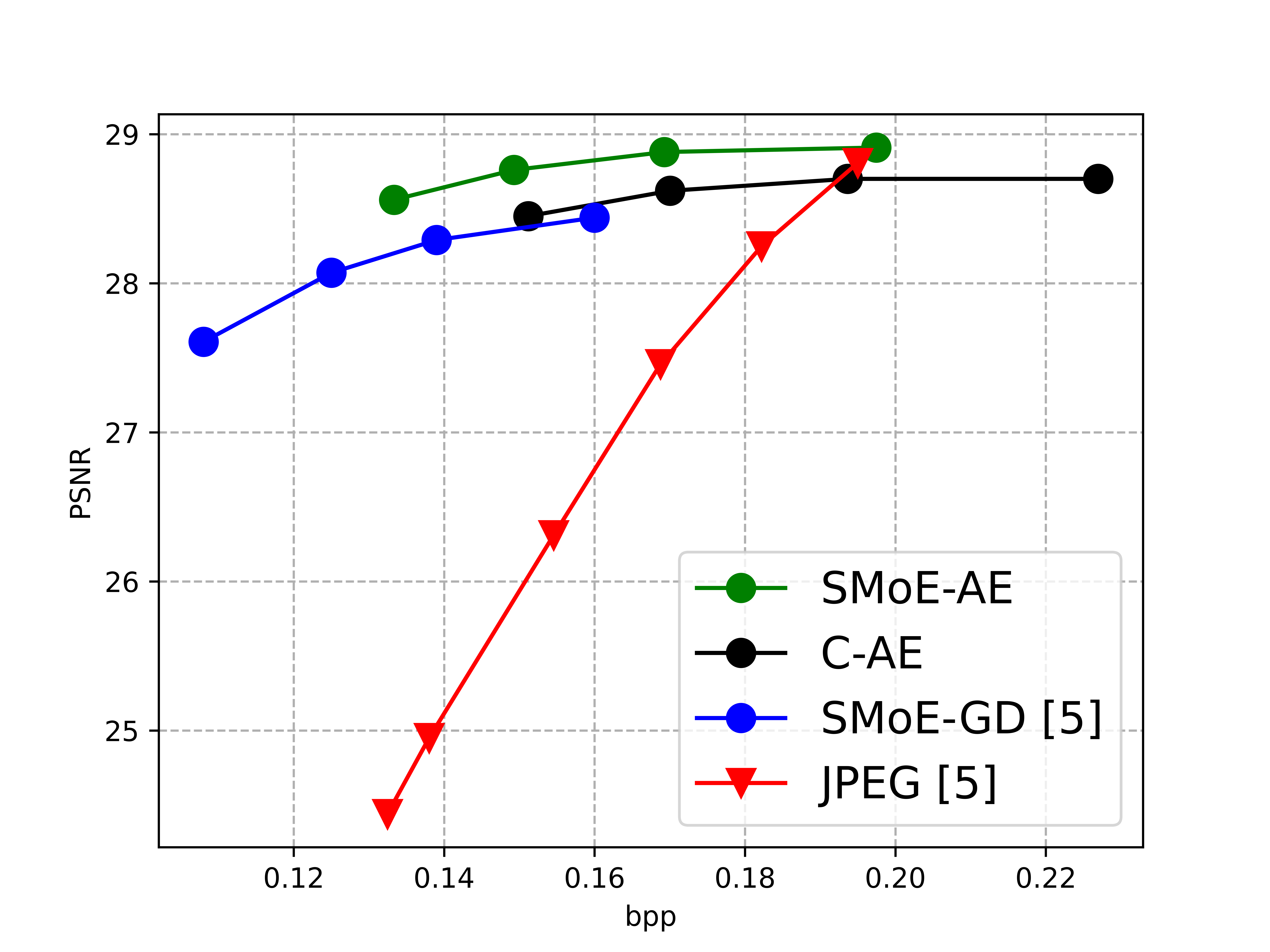}

\label{fig:Peppers_RD}
\end{subfigure}
\centering
\begin{subfigure}[t]{0.49\linewidth}
\includegraphics[width=\linewidth]{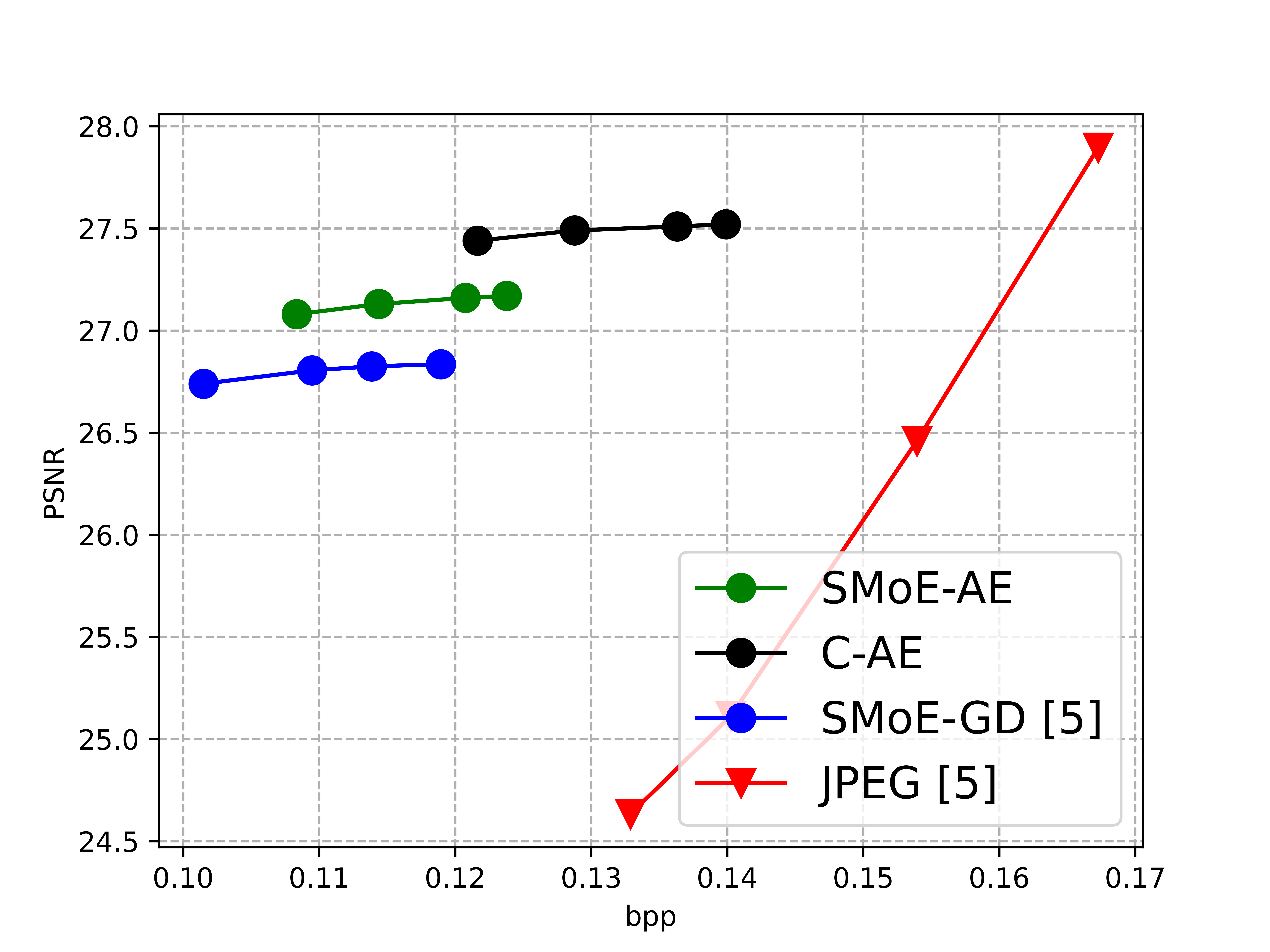}

\label{fig:Cameraman_RD}
\end{subfigure}%

\caption{RD-curves for Peppers (left) and Cameraman (right)}
\label{fig:RD_Curves}
\vspace{-0.2cm}
\end{figure}

The prime purpose was to understand trade-offs between image quality and run-time properties of the SMoE-AE approach compared to the complex SMoE-GD in \cite{tok2018mse} and C-AE. All SMoE experiments used the fixed number of four kernels and a bandwidth $S=0.0035$.\\
With known input pixels $y_p$ and the estimated gating $w_i$ the experts $m_i$ of Eq. \ref{eq:Grundgleichung} can be optimally recalculated by means of Ordinary Least Squares (OLS).
The estimation of these experts is still beneficial for the encoder training as it regularizes the network towards finding better center positions. Ommiting the expert estimation and solely relying on the optimal experts obtained by above mentioned OLS would degrade the overall performance by 0.1 dB in PSNR. \looseness=-1\\
Tab. \ref{tab:Tok_Time} shows a comparison of the reconstruction quality of SMoE-GD to the novel SMoE-AE network for the unquantized model output.
The results are based on $512\times512$ pixel gray-level images. 
For comparison, the reconstruction quality of the conventional C-AE Autoencoder is presented as a baseline.
The reconstruction quality of the of SMoE-GD depends on the number of training iterations during encoding while the run-time of the SMoE-AE network does not change as the SMoE parameters are estimated in a single pass.
Most noticeably: our SMoE-AE network consistently outperforms the SMoE-GD approach in \cite{tok2018mse} in terms of PSNR with comparable SSIM while reducing the run-time drastically, as seen in Tab. \ref{tab:Encode-Decode_Times}.\\
%
On average the SMoE-GD approach converges to an acceptable quality between 2500 - 5000 iterations, consuming between 150-300 seconds for each $512\times512$ image for encoding. This is brought down to 0.25 seconds/image with the novel SMoE-AE encoder network making the SMoE approach with SMoE-AE parameter prediction feasible for practical implementations. 
This results in a speed up gain by a factor of 500 to 1000 which we believe is remarkable given the even increased quality.
Even with 20.000 iterations the quality of SMoE-GD is still inferior to SMoE-AE in terms of PSNR and on par in SSIM. In the case of the Cameraman test image, the SMoE-GD optimization does not even seem to converge to a satisfying quality, which is due to initialization problems of the parameters. As outlined above, the SMoE-AE network does not need to initialize the SMoE model parameters and seems to deal with such situations robustly. Surprisingly, the C-AE Autoencoder does not necessarily outperform the SMoE-AE in terms of PSNR or SSIM. We found that on average over many test images the results appear comparable. Often it seems that distinct edges in images are better reconstructed using the edge-aware SMoE-AE approach as expected. An example can be seen in Fig.\ref{fig:AE_Smoe_Ausschnitte}.\\
It is further beneficial to investigate implementation properties of the decoders. Since SMoE-AE and C-AE encoders have identical encoder networks and generate identical numbers of bottleneck parameters (even though with different encoder network parameters) they have identical encoder run-time properties. The decoders, however, differ from each other. SMoE-GD and SMoE-AE decoders are identical with shallow network structure and naturally fast to compute. 
This accounts to 0.02 sec/image as seen in Tab. \ref{tab:Encode-Decode_Times}. On the other hand the C-AE decoder run-time is identical to its encoder run-time due to the symmetry of the AE and consumes 0.25 sec/image which is significantly larger compared to SMoE decoder.\\
Fig. \ref{fig:Comp_1_iteration} shows a visual comparison of the reconstruction quality of SMoE-GD for test image Peppers reported in Tab. \ref{tab:Tok_Time} to SMoE-AE and C-AE networks.
%
The PSNR of fully optimized SMoE-GD, SMoE-AE and C-AE are almost same but close inspection reveals that SMoE-GD and SMoE-AE provide in many cases significantly better edge quality. We attribute this to the inherent edge-awareness of the SMoE model.\\
Tab. \ref{tab:Comp_JPEG_Tok_AE} shows a comparison of the compression results between JPEG, SMoE-GD, SMoE-AE and C-AE at very low bit rates. To be comparable to JPEG the SMoE and C-AE parameters are quantized and the division into ”non- textured” and textured blocks is used identical to the work in \cite{tok2018mse}. For textured blocks, the SMoE center positions $\underline{\mu}_i$ are quantized with 3-4 bits/component and the expert values $m_i$ with 4-5 bits as in \cite{tok2018mse}. The average luminance value of non-textured blocks is coded with 8 bits. 
To distinguish between the block types one bit is added. 
For C-AE the 12 parameters of the bottleneck-layer are quantized with 4-5 bits. Note, that these parameters are now not predicting kernel positions and experts anymore and thus do not bear physical interpretation regarding edges etc. as with SMoE-GD and SMoE-AE.\\
Under quantization, the improvements reported in Tab. \ref{tab:Tok_Time} are preserved, as depicted in the Rate-Distortion curves in Fig. \ref{fig:RD_Curves} - maintaining the reported tremendous run-time savings of SMoE-AE. The above mentioned recalculation of the experts becomes more important when adapting to center positions after quantization and provides gains of up to 1dB in PSNR. Visual results for test image “Peppers” in Fig. \ref{fig:JPEG_SMOE} (enlarge pdf to see details in images) reveal that both SMoE-GD and SMoE-AE significantly outperform JPEG at very low bit rates, with excellent edge reconstruction properties. Also, SMoE-AE reconstruction is slightly better than quantized SMoE-GD, even though visually hardly noticeable.
The significant result is, however, that the novel SMoE-AE network provides a quantum leap in run-time improvement for SMoE compression approaches.

\begin{table*}
\begin{scriptsize}
\caption{Comparison JPEG, SMoE-GD, SMoE-AE and C-AE after quantization}
\centering
\begin{tabular}[c]{l l l l l l l l l l l l l }
\toprule
& \multicolumn{3}{c}{JPEG}& \multicolumn{3}{c}{SMoE-GD}&\multicolumn{3}{c}{SMoE-AE}&\multicolumn{3}{c}{C-AE}\\
\midrule
Sequence&Rate [bpp]&PSNR [dB]&SSIM &Rate [bpp]&PSNR [dB]&SSIM&Rate [bpp]&PSNR [dB]&SSIM&Rate [bpp]&PSNR [dB]&SSIM\\
\multirow{2}{*}{Cameraman}&0.15&26.45&0.73&0.08&26.47&\textbf{\color{gray}0.80}&0.08&26.69&0.79&0.08&\textbf{\color{gray}26.76}&0.78\\
&0.17&27.88&0.81&0.09&26.30&\textbf{\color{gray}0.81}&0.09&26.88&0.80&0.09&\textbf{\color{gray}27.04}&0.79\\
\multirow{2}{*}{Lena}&0.14&24.82&0.67&0.14&27.95&0.77&0.14&\textbf{28.30}&\textbf{0.78}&0.14&27.94&0.76\\
&0.17&27.32&0.74&0.17&28.11&0.78&0.17&\textbf{28.46}&\textbf{0.80}&0.17&28.20&0.78\\
\multirow{2}{*}{Peppers}&0.14&24.95&0.62&0.14&28.29&0.71&0.14&\textbf{28.69}&\textbf{0.75}&0.14&28.33&0.73\\
&0.17&27.46&0.69&0.17&28.46&0.75&0.17&\textbf{28.88}&\textbf{0.76}&0.17&28.62&0.74\\
\bottomrule
\end{tabular}

\label{tab:Comp_JPEG_Tok_AE}
\end{scriptsize}
\vspace{-0.2cm}
\end{table*}



\section{Summary/Conclusion}
In this paper we proposed an Autoencoder approach for training and coding Steered-Mixture-of-Experts models with the goal of speeding up encoding times for real-time applications, including compression. We emphasize, that such strategy has similarities and is in the spirit of "unfolding“ approaches currently being discussed with Deep Neural Networks \cite{hershey2014deep}\cite{zhang2020deep}.
We conclude that the SMoE-AE "unfolding" approach drastically reduces encoder run-time compared to the fully optimized SMoE-GD models using gradient-descent algorithms and very significantly the decoder run-time compared to C-AE conventional Autoencoders – without reducing reconstruction quality.
In contrast to C-AE, the “fast” SMoE-AE decoder preserves the high interpretability of the SMoE-GD parameters and the embedded edge-model.
Very importantly, SMoE-AE coding allows the additional initially-mentioned functionalities at the decoder, such the super-resolution-ready reconstruction, by resampling Eq. \ref{eq:OPGleichung} to any desired resolution.
By calculating gradients from Eq. \ref{eq:OPGleichung} (or analysing kernel positions in a block) edge and correlation information is easily accessible for further analysis/postprocessing at the decoder.\\
We believe our preliminary experiments and results comparing SMoE-AE with C-AE are not sufficient to explain the surprisingly good performance embedding the SMoE model into SMoE-AE. After all it seems counter-intuitive, that an Autoencoder with a shallow decoder is comparable if not better in reconstruction quality compared to a conventional Autoencoder. Both strategies were implemented with a standard encoder network and trained without “belts and whistles” that may or may not have improved performance against each other. A possible hypothesis is that the shallow, edge aware SMoE decoder fixed in size and parameters regularizes the optimization of the SMoE encoder– and/or otherwise leaves less parameters to optimize in the training process compared to C-AE which improves performance.\\
With the contributions in this paper the SMoE compression approach introduced in \cite{tok2018mse} is ready for real-time applications. The coding structure in its present form, with fixed number of 4 non-steering kernels/block, is however not competitive with state-of-the-art JPEG2000 or HEVC-Intra coding at higher bit rates. Our goal is to make the approach block-adaptive by employing variable number of kernels/block and/or blocks of different sizes, including steering kernels. To this end we will investigate more complex SMoE-AE encoder and decoder networks for advanced kernel parameter prediction.

\bibliographystyle{ieeetr}          
\bibliography{bibliography}

\end{document}